\pdfoutput=1

\documentclass[11pt]{article}

\usepackage[preprint]{acl}

\usepackage{times}
\usepackage{latexsym}
\usepackage{multirow}
\usepackage{booktabs}
\usepackage{tablefootnote}
\usepackage{arydshln}
\usepackage[normalem]{ulem}
\usepackage{graphicx}
\usepackage{listings}
\usepackage{colortbl}
\usepackage[T1]{fontenc}

\usepackage[utf8]{inputenc}
\usepackage{xspace}

\usepackage{microtype}

\usepackage{inconsolata}
\usepackage{algorithm}
\usepackage{algorithmic}
\usepackage{graphicx}
\usepackage{amsmath}
\usepackage{amssymb}
\usepackage{pgfplots}
\pgfplotsset{compat=1.18} 

\newcommand{\method}{\textsc{Eh-MAM}\xspace}

\title{\method: Easy-to-Hard Masked Acoustic Modeling for Self-Supervised Speech Representation Learning}

\author{
    Ashish Seth$^{1* \dagger}$,
    Ramaneswaran Selvakumar$^{1*}$,
    S Sakshi$^{1}$,
    Sonal Kumar$^{1}$,\\
    \bf Sreyan Ghosh$^{1 \dagger}$,
    \bf Dinesh Manocha$^{1}$ \\
    $^{1}$University of Maryland, College Park\\
    \texttt{\{aseth125, ramans, ssakshi, sonalkum, sreyang, dmanocha\}@umd.edu}
}

\begin{document}
\maketitle
\begingroup
\renewcommand\thefootnote{}
\footnotetext{* Equal contribution, $\dagger$ Equal Ideation}
\endgroup
\begin{abstract}


In this paper, we present \method (Easy-to-Hard adaptive Masked Acoustic Modeling), a novel self-supervised learning approach for speech representation learning. In contrast to the prior methods that use random masking schemes for Masked Acoustic Modeling (MAM), we introduce a novel selective and adaptive masking strategy. Specifically, during SSL training, we progressively introduce \textit{harder} regions to the model for reconstruction. Our approach automatically selects hard regions and is built on the observation that the reconstruction loss of individual frames in MAM can provide natural signals to judge the difficulty of solving the MAM pre-text task for that frame. To identify these hard regions, we employ a teacher model that first predicts the frame-wise losses and then decides which frames to mask. By learning to create challenging problems, such as identifying harder frames and solving them simultaneously, the model is able to learn more effective representations and thereby acquire a more comprehensive understanding of the speech. Quantitatively, \method outperforms several state-of-the-art baselines across various low-resource speech recognition and SUPERB benchmarks by 5\%-10\%. Additionally, we conduct a thorough analysis to show that the regions masked by \method effectively capture useful context across speech frames~\footnote{Code: \url{https://github.com/cs20s030/ehmam.git}}.


\end{abstract}

\everypar{\looseness=-1}
\section{Introduction}
Self-supervised learning (SSL) has emerged as one of the most effective paradigms of speech representation learning when labeled data is scarce~\cite{baevski2020effectiveness, mohamed2022self}. The task is to learn general-purpose speech representations from unlabeled data that can then be transferred to Spoken Language Processing (SLP) tasks like Automatic Speech Recognition (ASR), Speech Emotion Recognition (SER), etc~\cite{huang2001spoken}. Progress in SSL for speech has led to significant performance improvements in a range of low-resource SLP tasks including Phoneme Recognition (PR), Keyword Spotting (KS), etc~\cite{mohamed2022self}. Masked Acoustic Modeling (MAM) has been one the most prevalent pretext tasks for SSL-based speech representation learning wherein the model tries to reconstruct frames that are masked at the input, utilizing the context of the surrounding frames~\cite{baevski2022data2vec,baevski2023efficient}.

\begin{figure}
    \centering
    \includegraphics[width=\columnwidth]{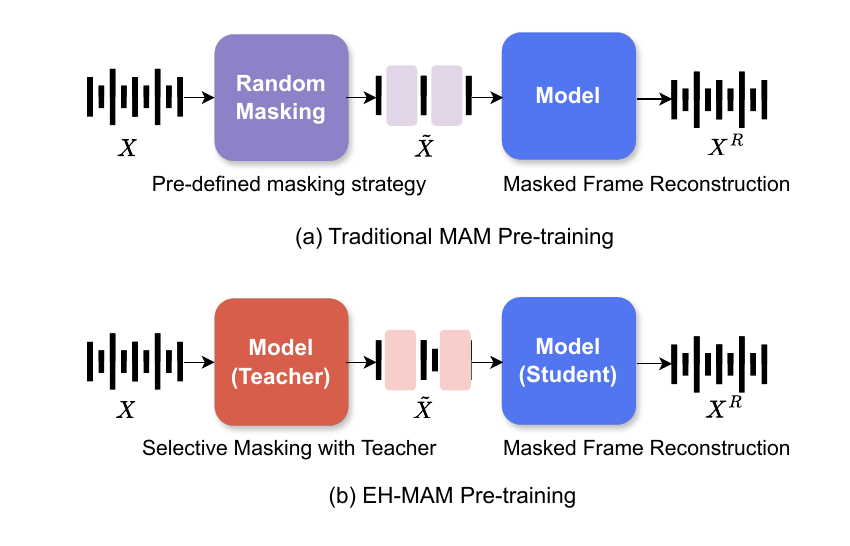}
    \caption{\small \method compared to random masking schemes employed widely in the literature. \method first identifies which frames to mask using a Teacher model and then solves the MAM task by reconstructing the selected masked regions using a Student model.}
    \label{fig:enter-label}
\end{figure}




\begin{figure}
    \raggedright
    \begin{tikzpicture}
        \begin{axis}[
            xlabel={Masking Percentage (\%) $\rightarrow$},
            ylabel={Relative WER $\rightarrow$},
            legend pos=north east,
            grid=major,
            width=\columnwidth,
            height=0.8\columnwidth,
            ymin=0, ymax=2.4,
            xtick={10,20,30,40,50},
            ytick={0.5,1.0,1.5,2.0,2.5},
            legend style={font=\small},
            xlabel style={font=\small},
            ylabel style={font=\small},
        ]
        
        \addplot[
            color=blue,
            mark=square,
            ]
            coordinates {
            (10, 0.48) (20, 0.69) (30, 0.8) (40, 1.0) (50, 1.21)
            };
        \addlegendentry{Random Masking}
        
        \addplot[
            color=red,
            mark=o,
            ]
            coordinates {
            (10, 0.96) (20, 1.13) (30, 1.19) (40, 1.31) (50, 1.48)
            };
        \addlegendentry{Selective Masking}

        \end{axis}
    \end{tikzpicture}
    \caption{\small Increase in relative WER using selective and random masking schemes. During inference, under similar experimental settings, we selectively mask the frames with high reconstruction values and compare it against random masking. The former consistently shows a significant increase in relative WER than the later, thereby indicating that these frames capture more useful context for speech reconstruction as a result of capturing more information, Thus building on this result we hypothesize that asking a model to reconstruct these frames will result in stronger learning signals.}
    \label{fig:wer_comparison}
\end{figure}
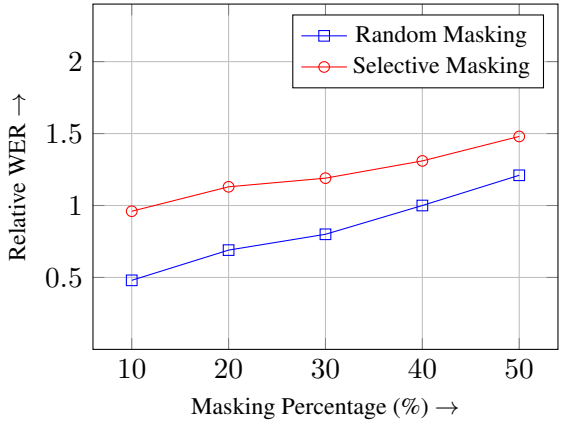

Although a considerable amount of research in MAM has been performed, most has focused on improving model architectures~\cite{baevski2022data2vec,chang2022distilhubert,baevski2023efficient} and pretext tasks~\cite{hsu2021hubert,lodagala2023data2vec,liu2024dinosr}, with very limited progress in improving the masking algorithm~\cite{yue2022self,baevski2023efficient}. Most MAM algorithms still perform random masking of input frames. On the other hand, selective masking strategies for other domains, like computer vision (CV)~\cite{bao2021beit,he2022masked,kakogeorgiou2022hide} and natural language processing (NLP)~\cite{sadeq2022informask,sadeq2023unsupervised,xiao2022retromae} that focuses on masking useful context, have shown significant improvements over random masking. This can be attributed to multiple factors, including: \emph{(1) Variable Information Content:} Variable information content in data translates to variable learning signals for the reconstruction task. For instance, in Masked Language Modeling (MLM)~\cite{devlin-etal-2019-bert}, the reconstruction of high-frequency stop words such as ``the'' or ``is'' offers minimal discriminative power due to the ubiquity and low semantic load of these words~\cite{sadeq2022informask, sadeq2023unsupervised}. In speech, for example, this can be translated to reconstructing frames corresponding to random noise or partial phonemes, where much of the frames is already available as context. \emph{(2) Progressive Learning:} Random masking fails to imitate the progressive human learning process~\cite{madan2024cl}. Humans do not receive knowledge uniformly; instead, they are exposed to progressively more complex information as they advance in the learning process. Mimicking this progression in the masking algorithm by initially exposing the model to simpler, more predictable speech patterns and gradually introducing more complex, less predictable ones can significantly enhance the learning trajectory. This approach aligns better with how humans learn, moving from simpler to more complex information, and helps the model develop a deeper understanding of language over time.



\vspace{0.1mm}
{\noindent \textbf{Main Contributions.}} To overcome the aforementioned problems, in this paper, we propose \method (\textbf{E}asy-To-\textbf{H}ard adaptive \textbf{M}asked \textbf{A}coustic \textbf{M}odelling), a novel selective and adaptive masking scheme for MAM.  We build \method on the core hypothesis that \emph{hard regions, characterized by collections of speech frames that are more difficult to reconstruct, serve as stronger signals for the learning process}. Fig.~\ref{fig:wer_comparison} shows the results of a simple experiment we performed to validate our hypothesis. By selectively masking \textit{hard} regions, we notice a greater and consistent drop in WER performance for ASR. This suggests that masking hard regions captures useful context in the speech input. Our main contributions are as follows:






\begin{figure*}[t]
    \centering
    \includegraphics[width=\linewidth]{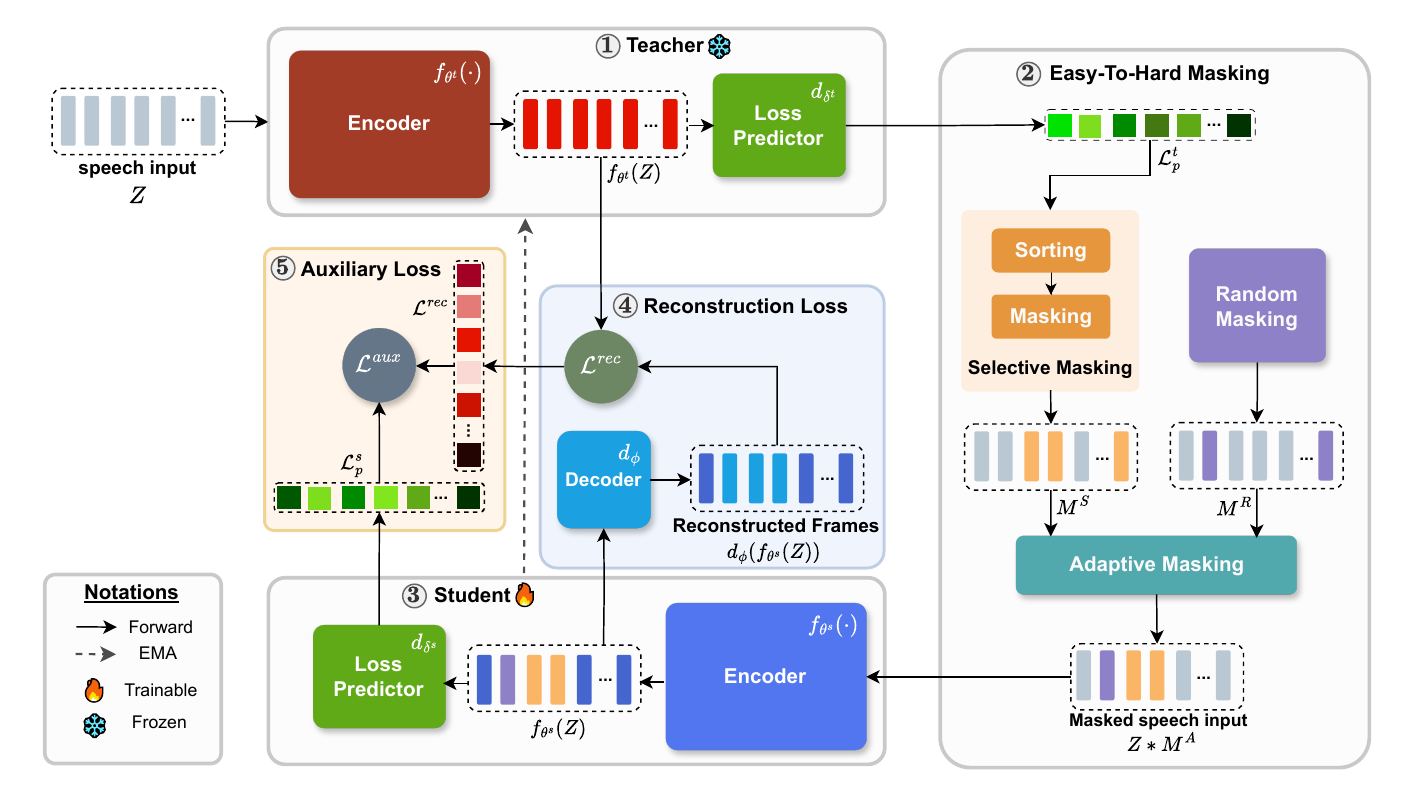}
    \caption{\small  Illustration of \method SSL algorithm. \method employs the self-distillation SSL framework that consists of identical student and teacher networks. At each training iteration, the teacher is updated by the exponential moving average (EMA) of the student. \textcircled{\raisebox{-0.9pt}{1}} For a speech input $Z$, we first use the teacher network to identify the speech frames that are hard to reconstruct, also called as hard regions. To achieve this, we predict the frame-level reconstruction loss values $\mathcal{L}^t_p$ using a loss predictor $d_{\delta^t}$ by feeding $Z$ to the teacher network. \textcircled{\raisebox{-0.9pt}{2}} Next, we utilize our \emph{easy-to-hard} masking strategy to identify the mask indices $M^S$ associated with hard regions, followed by progressively introducing them with random mask indices $M^R$ over each epoch. \textcircled{\raisebox{-0.9pt}{3}} Finally, a masked variant $\tilde Z$ is fed to the student network, where it is tasked to \textcircled{\raisebox{-0.9pt}{4}} reconstruct masked regions by optimizing a reconstruction loss (as shown in Eqtn.~\ref{eq:rec}) and \textcircled{\raisebox{-0.9pt}{5}} train a loss predictor $d_{\delta^s}$ by computing an auxiliary loss between predicted and original reconstruction loss values, $\mathcal{L}^s_p$ and $\mathcal{L}^{rec}$ respectively (as shown in Eqtn.~\ref{eq:aux}).} 
    \label{fig:eh_mam-arch}
\end{figure*}

\begin{itemize}
    \item We propose \method, a novel self-supervised speech representation learning algorithm. In contrast to solving a predefined MAM pre-text task, such as reconstructing randomly masked frames, \method aims to generate and align itself towards a more formidable MAM pre-text task. For generating a challenging MAM pre-text task, we first identify a collection of difficult frames to reconstruct, also called \emph{hard regions}, followed by selectively masking them. We propose a lightweight \emph{loss predictor} (introduced in Section~\ref{subsec:auxiliary}) that predicts the frame-level reconstruction loss values and determines \emph{hard regions} based on the output. To train the loss predictor jointly with MAM, we design a novel \emph{auxiliary loss} (introduced in Section~\ref{subsec:auxiliary}) that forces the predictor to learn the relative correlations between speech frames. Finally, to align the model towards reconstructing hard regions, we propose an \emph{easy-to-hard} masking strategy (introduced in Section~\ref{subsec:e2h}) that guides the \method learning. 
    \item We show the effectiveness of the speech representation learned by \method through extensive evaluations on low-resource speech recognition benchmarks~\cite{kahn2020libri} and downstream evaluation on SUPERB~\cite{yang2021superb}. \method beats prior arts with a relative improvement of 5\%-10\%
    \item We perform a comprehensive analysis to demonstrate that regions masked by the \method effectively capture useful context across speech input.
\end{itemize}





\section{Related Work}
\noindent{\textbf{Self-Supervised Learning.}} SSL has emerged as a prevalent speech representation learning paradigm, demonstrating impressive downstream performance under low-resource settings~\cite{lee-etal-2022-self, mohamed2022self}. At its core, SSL relies on the quality of pretext tasks for capturing varied learning signals from unlabeled data sources. Based on the nature of the pretext tasks, the SSL frameworks are further categorized into the following sub-categories: 1) Contrastive Approaches: The pretext task is designed to maximize latent space similarity between the anchor and positive samples while minimizing the similarity between the anchor and negative samples. 2) Generative Approaches: These methods primarily focus on first building a target by randomly masking multiple speech frames and then reconstructing them by optimizing a similarity measure (MSE or Cross Entropy) between the predicted frames and the targets. The pretext task includes predicting future input from past inputs~\cite{oord2018representation, yang2022autoregressive}, masked from unmasked~\cite{baevski2022data2vec, baevski2023efficient} or the original from some other corrupted view~\cite{lodagala2023data2vec}. Masked Acoustic Modeling has undoubtedly seen the most success for speech representation learning.



\noindent{\textbf{Masked Acoustic Modeling (MAM)}}
Conventional MAM architectures first perform frame-level masking, where randomly selected speech frames are masked using various existing masking strategies, including block or random masking~\cite{bao2021beit, he2022masked}. Next, they either employ a single encoder network like BERT~\cite{devlin-etal-2019-bert} to predict masked regions in a speech input~\cite{9054458,chang2022distilhubert,chen2022wavlm,hsu2021hubert} or utilize self-distillation methods, where the student learns to reconstruct masked information under the guidance of an identical teacher network~\cite{baevski2022data2vec,baevski2023efficient,liu2024dinosr}. Although a considerable amount of research in MAM has undergone towards improving model architecture~\cite{baevski2022data2vec,baevski2023efficient} and introducing novel pretext tasks~\cite{liu2024dinosr}, developing better masking strategies is still under-explored.

\section{Methodology}
In this Section, we explain the \method methodology. We first provide an overview of the \method learning paradigm (in Section~\ref{sec:method_overview}), followed by details on the reconstruction and auxiliary loss formulations (in Sections~\ref{subsec:reconst},~\ref{subsec:auxiliary}). Finally, we introduce the \emph{easy-to-hard} masking algorithm (in Section~\ref{subsec:e2h}).

\subsection{Overview of \method}
\label{sec:method_overview}
We illustrate \method in Fig.~\ref{fig:eh_mam-arch}. At its core, \method incorporates the \emph{self-distillation} based SSL training paradigm for solving MAM pretext task, similar to~\citet{baevski2022data2vec,baevski2023efficient}. Specifically, \method consists of two identical networks, a teacher $\{f_{\theta^t}, d_{\delta^t}\}$ and a student $\{f_{\theta^s}, d_{\delta^s}\}$. A separate decoder $d^{R}_{\phi}$ is employed for reconstructing masked frames from the student representations. The context encoders $f_{\theta}$ are built using $K$-layered transformers~\cite{vaswani2017attention}, whereas the decoder $d^{R}_{\phi}$ and the loss predictor $d_{\delta}$ are constructed with light-weight $D$-layered 1D-convolution layers~\cite{kiranyaz20191}. During pre-training, the teacher parameters ${\theta^t,\delta^t}$ are updated by performing exponential moving average (EMA) of the student parameters ${\theta^s,\delta^s}$~\cite{tarvainen2017mean}. Formally, we define the update as follows:
\begin{equation}
    \omega^t = \lambda \omega^t + (1-\lambda) \omega^s
\end{equation}
where $\omega^t = \{\theta^t, \delta^t\}$, $\omega^s = \{\theta^s, \delta^s\}$, and $\lambda$ is the decay rate. The student and decoder parameters are updated using gradient descent.

At each training iteration, we first extract low-frequency feature representations $Z \in \mathbb{R}^{N \times d}$ from raw speech signals $x \in X$~\cite{baevski2020wav2vec} and feed it to the teacher network to get frame-level predicted reconstruction loss values $\mathcal{L}^t_p = d_{\delta^t}(f_{\theta^t}(Z))$. With the help of $\mathcal{L}^t_p$, we generate binary mask indexes $M^A=\{0,1\}^{N}$ using the \emph{easy-to-hard} masking strategy (introduced in Section~\ref{subsec:e2h}), followed by creating a masked version of the original speech input $\tilde Z \leftarrow Z \cdot M^A$. Finally, the student is trained with gradient descent to minimize a weighted combination of reconstruction loss $\mathcal{L}^{rec}$ (introduced in Section~\ref{subsec:reconst}) and an auxiliary loss $\mathcal{L}^{aux}$ (introduced in Section~\ref{subsec:auxiliary}). Formally, we define the objective function below: 
\begin{equation}
    \mathcal{L}^{joint} = \mathcal{L}^{rec} + \alpha \mathcal{L}^{aux}
\end{equation}
where $\alpha$ is a balancing parameter and is set to $0.05$ throughout the experiments (further ablation on this can be found in Appendix~\ref{subsec:bal_par}).

\begin{algorithm}[t]
\caption{Easy-To-Hard Masking}
\label{alg:easy-to-hard}
\begin{algorithmic}[]
\REQUIRE train set $\mathcal{Z}$, masking probability $\mathcal{P}$, selective masking $M^S$, random masking $M^R$, adaptive masking $M^A$, number of frames $F(\cdot)$  
\REQUIRE $t \in \{1,2,...,T\}$: training iteration
\FOR{$t \le T$}
    \STATE Sample batch of training data $z \sim \mathcal{Z}$
    \STATE Compute selective and random masking probability: $\mathcal{P}^S \leftarrow \mathcal{P}\times\frac{t}{T}$, $\mathcal{P}^R \leftarrow 1-\mathcal{P}^S$ respectively.
    \STATE Predict the reconstruction value $\mathcal{L}^t_p$ for each frame: $\mathcal{L}^t_p \leftarrow d^T_{\delta}(f_{\theta^t}(z))$ 
    \STATE Update $M^S$ by selecting frame indices corresponding to the top $k$ reconstruction values where $k=\lfloor P^S F(z) \rfloor$. 
    \STATE Update $M^R$ by randomly selecting $\lfloor P^R F(z) \rfloor$ frame indices
    \STATE Update $M^A$ by taking the union of $M^S$ and $M^R$: $M^A \leftarrow M^S \cup M^R$
    \STATE Create a masked counterpart $\tilde z \leftarrow M^A \cdot z$
    
\ENDFOR
\end{algorithmic}
\end{algorithm}
\subsection{Selective Masking with \method}
\label{subsec:selective_masking}
\noindent{\textbf{Motivation:}} \method distinguishes itself from the conventional self-distillation-based SSL training methods that are fixated on solving a predefined MAM task generated using random masking~\cite{baevski2022data2vec,chen2022wavlm,chang2022distilhubert}, by enforcing the teacher to generate more challenging pretext tasks. To achieve this, \method first uses the teacher to identify hard regions, a collection of speech frames that are difficult to reconstruct, and then selectively mask these hard regions to create challenging MAM pretext tasks for the student to solve. Being constantly challenged by the teacher further directs the student to develop a much more nuanced understanding of speech. Additionally, we take inspiration from the recent studies in NLP and CV that have highlighted the significance of generating formidable pretext tasks for MLM (Masked Language Modeling) and MIM (Masked Image Modeling) using selective masking~\cite{bao2021beit,sadeq-etal-2022-informask}.  

To reweigh the model attention towards reconstructing such hard regions, we introduce the loss predictors $d_{\delta^s}$, $d_{\delta^t}$ for the student and teacher networks, respectively. Further to train the loss predictor, we also propose an auxiliary objective function $\mathcal{L}^{aux}$, that the model optimizes alongside the reconstruction loss $\mathcal{L}^{rec}$.

\subsubsection{Reconstruction Loss} 
\label{subsec:reconst}
As shown in Fig.~\ref{fig:eh_mam-arch}, we first reconstruct the masked frames by feeding student representations $f_{\theta^s}(\tilde Z)$ to a decoder $d^{R}_{\phi}$. Similar to~\citet{baevski2022data2vec,baevski2023efficient}, the goal of $d^{R}_{\phi}$ is to reconstruct the teacher representation for time steps that are masked in the student input. To achieve this, we compute a reconstruction loss $\mathcal{L}^{rec}$ between the student and the teacher representations. Formally, we define reconstruction loss $\mathcal{L}^{rec}$ as follows:
\begin{equation}
    \mathcal{L}^{rec} = \| M^A \cdot f_{\theta^t}(Z) - d^{R}_{\phi}(\cdot f_{\theta^s}(\tilde Z)) \|_2^2
\label{eq:rec}    
\end{equation}
where $M^A \cdot f_{\theta^t}(Z)$ represents teacher representations associated with the masked speech input.

\subsubsection{Loss Predictor and Auxiliary Loss}
\label{subsec:auxiliary}
\noindent{\textbf{Motivation:}} Given the sequence of frame-level reconstruction loss values $\mathcal{L}^{rec} \in \mathbb{R}^{N}$, our goal is to create a challenging MAM pretext task for the student by selectively masking frames with high reconstruction values. As original reconstruction loss values $\mathcal{L}^{rec}$ are computed only for the masked regions (see Section~\ref{subsec:reconst}), it provides limited information for deciding which frames to mask. To mitigate this problem, we introduce lightweight loss predictors $d_{\delta^s}, d_{\delta^t}$, which can be easily integrated with the student-teacher network, and add reconstruction loss predicting capabilities across both networks. To train these loss predictors, we propose a novel auxiliary loss $\mathcal{L}^{aux}$ that guides it towards capturing relative correlations between individual frames rather than forcing the predictor to generate exact frame-level reconstruction values.

Specifically, for each masked frame $(i,j)$ where $i \neq j$ and $(i,j) \in \{1,2,...,N\}$, if $\mathcal{L}^{rec}_i > \mathcal{L}^{rec}_j$ than the predicted counterpart $\mathcal{L}^{s}_p = d_{\delta^s}(f_{\theta^s}(\tilde Z))$ must also have $\mathcal{L}^{s}_{p_i} > \mathcal{L}^{s}_{p_j}$. To formulate this constraint as a differentiable objective function, we first define a target distribution as an indicator variable $I$ that captures the relative correlations between original reconstruction loss values, such as $\mathcal{L}^{rec}_i > \mathcal{L}^{rec}_j$. Formally we define this as follows:
\begin{equation}
I_{i,j} = \begin{cases}1, & \mathcal{L}^{rec}_i>\mathcal{L}^{rec}_j \text { and } \{i,j\} \in M^A \\ 0, & \text { otherwise }\end{cases}
\label{eq:indicator}
\end{equation}   
Next, similar to $I$, we introduce a predicted distribution $S$ for representing the relative differences in the predicted reconstruction values $\mathcal{L}^{s}_p$. $S$ is formally defined with a $sigmoid$ function as:
\begin{equation}
     S_{i,j} = e^{(\mathcal{L}^{s}_{p_i}-\mathcal{L}^{s}_{p_j})}/(1+e^{(\mathcal{L}^{s}_{p_i}-\mathcal{L}^{s}_{p_j})})
\label{eq:sig}
\end{equation}
where $S_{i,j} > 0.5$ if $\mathcal{L}^{s}_{p_i} > \mathcal{L}^{s}_{p_j}$. Finally, we formulate our auxiliary objective function $\mathcal{L}^{aux}$ by first computing a vanilla cross entropy $\mathcal{H}(\cdot)$ between the target distribution $I$ and the predicted distribution $S$: $\mathcal{L}^{aux} \leftarrow \mathcal{H}(I,S)$ and then minimizing it jointly with the reconstruction loss. We define the formulation of $\mathcal{L}^{aux}$ below:

\begin{equation}
    \mathcal{L}^{aux} = -\sum^{N}_{i=1} \sum^{N}_{j=1} I_{i,j}\log S_{i,j} + \tilde I_{i,j}\log(1-S_{i,j})
\label{eq:aux}
\end{equation}
where $\tilde I_{i,j}=1-I_{i,j}$. $\{i,j\} \in M^A$ means that the $i$ and $j$ frames are masked during pre-training. 

\pgfplotstableread{
X Y Value
10 0 10.429948415660606
10 1 10.420834870388848
10 2 10.42781939962897
10 3 10.426917447594969
10 4 10.413729658596562
10 5 10.43032364570063
10 6 10.414813268406103
10 7 10.430296588610155
10 8 10.423029109752889
10 9 10.416423265409573
10 10 10.425968720502144
10 11 10.415839166048846
10 12 10.41968800356641
10 13 10.41579830043378
10 14 10.423534872357171
10 15 10.412638866292609
10 16 10.429412193875015
10 17 10.41841481704751
10 18 10.420225567143303
10 19 10.41727979866773
20 0 9.933915701194428
20 1 9.912892565191207
20 2 9.905914865329427
20 3 9.921198933194644
20 4 9.949270679993823
20 5 9.927570670156287
20 6 9.901595096755942
20 7 9.932884971832129
20 8 9.903151586124924
20 9 9.903288549430695
20 10 9.916703342183698
20 11 9.91610496433434
20 12 9.9258035275494
20 13 9.945845118508718
20 14 9.925972247949884
20 15 9.937187203005813
20 16 9.92515532711451
20 17 9.947443939128933
20 18 9.943059518579068
20 19 9.948080323009565
30 0 9.177002796374515
30 1 9.188510053361727
30 2 9.155973993167553
30 3 9.183401108923313
30 4 9.195264933910527
30 5 9.151204192361378
30 6 9.15151078992221
30 7 9.15148564189183
30 8 9.167600195657965
30 9 9.166363172237787
30 10 9.193126932663215
30 11 9.164393353516797
30 12 9.181367428219534
30 13 9.151649091333068
30 14 9.16434064852303
30 15 9.17431098756008
30 16 9.185743535111353
30 17 9.18422506113308
30 18 9.184781575550694
30 19 9.160210315236153
40 0 8.804862147262664
40 1 8.82661905294833
40 2 8.802164231505216
40 3 8.810052609076111
40 4 8.803109069208011
40 5 8.845003196802232
40 6 8.825186288026497
40 7 8.822506124128887
40 8 8.843590540616063
40 9 8.846420991508388
40 10 8.830747171096968
40 11 8.851967445154608
40 12 8.855782104687322
40 13 8.851364228319447
40 14 8.847757745630565
40 15 8.835634061815753
40 16 8.80378778782009
40 17 8.798186827269115
40 18 8.847798662256723
40 19 8.817185264465095
50 0 8.098092205272385
50 1 8.030326836045576
50 2 8.093351662240616
50 3 8.047215572625628
50 4 8.04502078220669
50 5 8.109809962572578
50 6 8.052723474227117
50 7 8.082124003083505
50 8 8.070958350884716
50 9 8.039270235909395
50 10 8.02777247322501
50 11 8.102229194104778
50 12 8.025990849383655
50 13 8.034451946182116
50 14 8.026321459174659
50 15 8.102106804174602
50 16 8.091632403172307
50 17 8.067065125963763
50 18 8.078191890657845
50 19 8.09185476412923
60 0 8.240618363078823
60 1 7.930367778974606
60 2 8.305595960360447
60 3 8.22707435410704
60 4 8.290733128745956
60 5 8.024121442746056
60 6 8.117531125661861
60 7 8.376000945052192
60 8 7.989714575456872
60 9 7.900431684292431
60 10 8.226538666449603
60 11 8.299391503946124
60 12 8.267178213664135
60 13 8.027019900249655
60 14 8.1912343458432
60 15 8.114476899781387
60 16 8.207770511994259
60 17 7.990622001527901
60 18 8.01875366485886
60 19 7.9876163364778145
70 0 6.701558427917138
70 1 6.870253116992866
70 2 6.960892753299753
70 3 6.391889637763262
70 4 7.019250175316362
70 5 6.687196467360513
70 6 6.637311103425712
70 7 6.87675926623531
70 8 6.630736661073949
70 9 6.676399382258121
70 10 7.075826236667002
70 11 6.340857206938718
70 12 7.073554718723344
70 13 6.8124113655071215
70 14 6.6810808847024985
70 15 6.332928759998537
70 16 6.639430071530288
70 17 6.96414114178656
70 18 6.488116669910625
70 19 6.597136422557788
80 0 5.819509842939718
80 1 5.879040243575568
80 2 5.963797912282725
80 3 6.135036095046636
80 4 6.678901429962777
80 5 6.179443177672506
80 6 6.651980657096077
80 7 6.807496405347683
80 8 6.865540606322938
80 9 6.425229544364433
80 10 6.291426299277645
80 11 6.378044403645062
80 12 5.910437504672508
80 13 6.3265305737299276
80 14 6.351467134658697
80 15 6.121518814442235
80 16 5.938498629931829
80 17 6.808308902082769
80 18 5.77080544131529
80 19 6.030458999185361
90 0 5.256652632893449
90 1 5.855176706829883
90 2 5.545112625787678
90 3 6.385124804057601
90 4 6.06700080939639
90 5 5.661213899410077
90 6 5.659318840177783
90 7 6.265413633445174
90 8 5.549676368860409
90 9 6.869829561653678
90 10 5.546199625537221
90 11 6.053257550433017
90 12 5.730285595467649
90 13 6.411934209281599
90 14 6.6501071895243635
90 15 5.378329753248458
90 16 5.624426643879172
90 17 5.4827921534089565
90 18 6.633241813661286
90 19 5.614896720760856
100 0 4.576133960788101
100 1 5.978183838573512
100 2 6.379731460163936
100 3 5.181925555527757
100 4 4.538341392287289
100 5 6.037142166590667
100 6 4.846513726940698
100 7 5.910242623126552
100 8 6.1401908461164245
100 9 6.082065089913199
100 10 5.386272100935225
100 11 4.780767428911284
100 12 5.548588504307174
100 13 5.34559244628171
100 14 6.239030577355949
100 15 5.3015725808476875
100 16 4.639545834804793
100 17 5.572694231265088
100 18 6.266187340579741
100 19 5.344643959941679
}\datatable

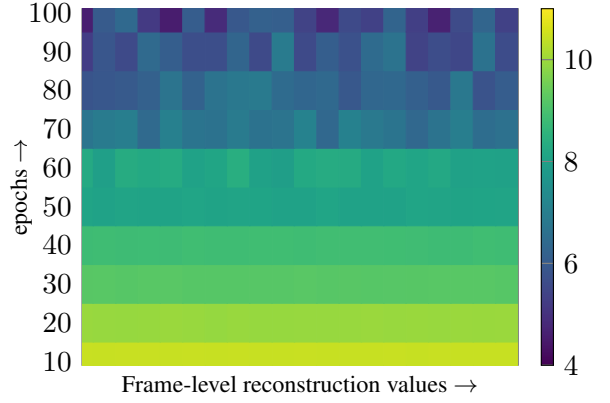
\begin{figure}
\raggedright

\begin{tikzpicture}
\begin{axis}[
    width=0.95\columnwidth,
    colorbar,
    colormap/viridis,
    colorbar style={
        width=0.01\textwidth, 
    },
    point meta min=4.0, 
    point meta max=11, 
    xmin=0, xmax=19.5, 
    ymin=9, ymax=101, 
    xtick=\empty, 
    ytick={10,20,30,40,50,60,70,80,90,100},
    ylabel=\small epochs $\rightarrow$,
    ylabel style={
        yshift=-9pt,
    },
    xlabel=\small Frame-level reconstruction values $\rightarrow$,
]
\addplot [
    matrix plot*,
    point meta=explicit,
    mesh/cols=20, 
] table [x=Y, y=X, meta=Value] {\datatable};
\end{axis}
\end{tikzpicture}
\caption{\small For a random speech utterance, we show the variation in frame-level reconstruction loss values across training epochs. During the initial stages of \method pre-training, we find that the model exhibits high frame-level reconstruction loss values, which results in low distinctiveness amongst individual values. This leads to increased stochasticity in the selective masking.}
\label{fig:recon}
\end{figure}
\subsubsection{Selecting Hard Regions for Reconstruction}
\label{subsec:e2h}

\noindent{\textbf{Motivation:}} Fig.~\ref{fig:recon} shows that during the initial stage of a \method pre-training, the reconstruction loss values are significantly high and exhibit low discriminative power ($L^{rec}_i \approx L^{rec}_j$). This leads to increased stochasticity in the overall selective masking process. Thus, inspired by the general human learning approach, where humans do not perceive knowledge uniformly but are subjected to a learning environment where they progressively comprehend more complex information, we propose an \emph{easy-to-hard} masking strategy that guides the model to progressively mask \textit{harder} regions for reconstruction. Specifically, we linearly increase the proportion of mask indices associated with hard regions at each training epoch. We define \textit{hard} regions as a collection of speech frames that the model finds difficult to reconstruct.

We illustrate the masking strategy in Fig.~\ref{fig:eh_mam-arch}. At each training iteration $t$ and with a masking percentage $P$, we first compute $P^S$ and $P^R$, the individual masking percentages for selective and random masking respectively. Precisely, we update $P^S$ and $P^R$ linearly as $P^S = P \times \frac{t}{T}$ and $P^R = 1 - P^S$, where $T$ is the total number of training iterations. In selective masking, for each sampled batch $z \in Z$, we build mask $M^S$ by selecting frame indices associated with the top $k$ predicted reconstruction values $\mathcal{L}^t_p$. We use $k = \lfloor P^S F(z) \rfloor$, where $F(z)$ denotes the number of speech frames for an input batch $z$. To build a random mask $M^R$, we randomly sample $\lfloor P^R F(z) \rfloor$ frame indices. Finally, we compute the adaptive mask $M^A$ by taking a union of $M^S$, $M^R$. We summarize the complete process of easy-to-hard masking in Algorithm ~\ref{alg:easy-to-hard} 


\section{Experimental Setup}
\label{sec:exp}
{\noindent{\textbf{Pre-training}}} Following~\citet{baevski2022data2vec, baevski2023efficient, liu2024dinosr}, we pre-trained our model with 960 hours of unlabelled speech from LibriSpeech corpus~\cite{panayotov2015librispeech}. Due to resource constraints, we use a base variant of the context encoder~\cite{baevski2020wav2vec}, with the number of transformer layers $K$ = 12 and masking percentage = 50\%. For the loss predictor and the reconstruction decoder, we utilize 1D-convolution layers, with the number of convolution layers $D$ = 4. Moreover, a balancing parameter $\alpha$ is introduced and set to 0.05 during the joint optimization of reconstruction and auxiliary loss. All the pre-training experiments are performed on 4 $\times$ A100 40GB GPUs, for 400k updates and using a batch size of 63 minutes of speech (Additional details on the hyper-parameters used in \method can be found in Table~\ref{tab:finetune_hparams}).
\vspace{0.5mm}

{\noindent{\textbf{Fine-tuning}}} Similar to~\citet{liu2024dinosr}, to show the effectiveness of the learned speech representation, we fine-tune only the student counterpart with an additional CTC layer~\cite{graves2006connectionist}. We conduct a comprehensive evaluation under a low-resource labeled data setting using a 10 mins / 1 hour / 10 hours split from LibriLight benchmark~\cite{kahn2020libri} and 100 hours split from Librispeech~\cite{panayotov2015librispeech}. For all the splits, we follow a similar fine-tuning setup as wav2vec2~\cite{baevski2020wav2vec} (We provide additional fine-tuning details for all the splits in the Appendix ~\ref{sec:asr_evaluation}). We also perform a SUPERB (Speech Processing Universal PERformance Benchmark) evaluation~\cite{yang2021superb}, where a separate prediction head is trained on top of a frozen pre-trained model for various downstream tasks (Additional details on the downstream tasks present in SUPERB can be found in Appendix~\ref{sec:superb}) 
\vspace{0.5mm}

{\noindent{\textbf{Baselines}}} We compare the performance of \method across various SSL-based speech representation learning baselines that employ 1) single encoder: wav2vec 2.0~\cite{baevski2020wav2vec}, HuBERT~\cite{hsu2021hubert} and 2) self-distillation network: data2vec~\cite{baevski2022data2vec}, data2vec 2.0~\cite{baevski2023efficient} and DinoSR~\cite{liu2024dinosr} to reconstruct masked frames (Additional details on all the baselines can be found in Appendix~\ref{sec:baseline_details}). Due to compute constraints, we avoid retraining the baselines from scratch and use the checkpoints open-sourced by the authors.
\vspace{0.5mm}

{\noindent{\textbf{Dataset and Evaluation Metric}}} We pre-train \method on 960 hours of unlabelled speech data from the LibriSpeech corpus~\cite{panayotov2015librispeech}. Further, we evaluate \method on a wide range of speech-related downstream tasks, including 1) Low resource ASR benchmarks: Libri-Light~\cite{kahn2020libri}, 100 hours LibriSpeech corpus~\cite{panayotov2015librispeech}, Wall-Street Journal (WSJ)~\cite{paul1992design}, SwitchBoard~\cite{godfrey1992switchboard} and 2) SUPERB evaluation: a collection of a diverse set of downstream tasks including Phoneme Recognition (PR), Automatic Speech Recognition (ASR), Keyword Spotting (KS), Intent Classification (IC) and Slot Filling (SF). Additional details on duration, train/test splits, and evaluation metrics can be found in Appendix~\ref{sec:dataset_details}.    

\begin{table}[t]
\resizebox{\columnwidth}{!}{
\begin{tabular}{ccccccc}
\toprule
\toprule
\multirow{3}{*}{Model} & \multicolumn{3}{c}{Content} & \multicolumn{3}{c}{Semantic}      \\ \cmidrule(l){2-7} 
                       & PR       & ASR     & KS      & IC       & \multicolumn{2}{c}{SF} \\ \cmidrule(l){2-7} 
                       & PER $\downarrow$    & WER $\downarrow$    & Acc $\uparrow$    & Acc $\uparrow$      & F1 $\uparrow$        & CER $\downarrow$       \\ \midrule
wav2vec 2.0            & 5.47    & 6.43    & 96.23   & 92.35    & 88.30      & 24.77     \\
HuBERT                 & 5.41    & 6.42    & 96.30   & \underline{98.34}    & 88.53      & \underline{25.20}     \\
WavLM                  & 4.84    & 6.31    & 96.79   & \textbf{98.63}    & \underline{89.38}      & 22.86     \\
data2vec               & 4.69    & 4.94    & 96.56   & 97.63    & 88.59      & 25.27     \\
DinoSR                 & \textbf{3.21}    & \textbf{4.71}    & \underline{96.89}   & 98.02    & 88.83      & 23.57     \\
data2vec 2.0           & 3.93    & 4.91    & \underline{96.89}   & 98.01    & 88.24      & 22.09     \\ \cdashline{1-7}
\method                   & \underline{3.86}    & \underline{4.89}    & \textbf{97.01}   & 98.01    & \textbf{89.47}      & \textbf{22.04}     \\ \bottomrule
\end{tabular}}
\caption{\small Results on Speech Processing Universal PERformance Benchmark (SUPERB). The downstream tasks include phoneme recognition (PR), automatic speech recognition (ASR), keyword spotting (KS), intent classification (IC), and slot filling (SF). The evaluation metrics used are accuracy (Acc), phoneme error rate (PER), word error rate (WER), f1 score (F1), and concept error rate (CER). The best and the second best results are \textbf{bolded} and \underline{underlined} respectively.}
\label{tab:superb_table}
\end{table}
\begin{table*}[!t]
  \centering
  \resizebox{0.8\textwidth}{!}{
  \begin{tabular}{lccccccc}
  \toprule
    \hline
    \multirow{2}{*}{Models} & \multirow{2}{*}{Pre-training steps} & \multirow{2}{*}{Batch size (minutes)} & \multicolumn{2}{c}{dev (WER $\downarrow$)} & \multicolumn{2}{c}{test (WER $\downarrow$)} \\
    \cmidrule{4-7}
     & & & clean & other & clean & other \\
    \midrule
    \multicolumn{7}{l}{\textbf{10 minutes labeled data}} \\
    wav2vec 2.0~\cite{baevski2020wav2vec} & 400k & 96 & 8.9 & 15.7 & 9.1 & 15.6 \\
    HuBERT~\cite{hsu2021hubert} & 250k + 400k & 47 & 9.1 & 15.0 & 9.7 & 15.3 \\
    data2vec~\cite{baevski2022data2vec} & 400k & 63 & 7.3 & 11.6 & 7.9 & 12.3 \\
    DinoSR~\cite{liu2024dinosr} & 400k & 63 & 6.6 & 10.8 & 7.3 & 11.8 \\
    data2vec 2.0~\cite{baevski2023efficient} & 400k & 17 & \underline{6.4} & \underline{10.5} & \underline{7.2} & \underline{11.5} \\ \cdashline{1-7}
    \method & 400k & 63 & \textbf{6.3} & \textbf{10.2} & \textbf{7.1} & \textbf{11.1} \\
    \midrule
    \multicolumn{7}{l}{\textbf{1 hr labeled data}} \\
    wav2vec 2.0~\cite{baevski2020wav2vec} & 400k & 96 & 5.0 & 10.8 & 5.5 & 11.3 \\
    HuBERT~\cite{hsu2021hubert} & 250k + 400k & 47 & 5.6 & 10.9 & 6.1 & 11.3 \\
    WavLM~\cite{chen2022wavlm} & 250k + 400k & 187 & - & - & 5.7 & 10.8 \\
    data2vec~\cite{baevski2022data2vec} & 400k & 63 & 4.0 & 8.5 & 4.6 & 9.1 \\
    DinoSR~\cite{liu2024dinosr} & 400k & 63 & 4.1 & 8.1 & 4.6 & 8.7 \\
    data2vec 2.0~\cite{baevski2023efficient} & 400k & 17 &  \underline{4.0} &  \underline{8.0} &  \underline{4.6} &  \underline{8.7} \\ \cdashline{1-7}
    \method & 400k & 63 & \textbf{4.0} & \textbf{7.8} & \textbf{4.6} & \textbf{8.7} \\
    \midrule
    \multicolumn{7}{l}{\textbf{10 hr labeled data}} \\
    wav2vec 2.0~\cite{baevski2020wav2vec} & 400k & 96 & 3.8 & 9.1 & 4.3 & 9.5 \\
    HuBERT~\cite{hsu2021hubert} & 250k + 400k & 47 & 3.9 & - & 4.3 & 9.4 \\
    WavLM~\cite{chen2022wavlm} & 250k + 400k & 187 & - & - & 4.3 & 9.2 \\
    data2vec~\cite{baevski2022data2vec} & 400k & 63 & 3.3 & 7.5 & 3.9 & 8.1 \\
    DinoSR~\cite{liu2024dinosr} & 400k & 63 & 3.1 & 7.0 & 3.6 & 7.6 \\
    data2vec 2.0~\cite{baevski2023efficient} & 400k & 17 & \underline{3.0} & \underline{7.0} & \underline{3.4} & \underline{7.6} \\ \cdashline{1-7}
    \method & 400k & 63 & \textbf{3.0} & \textbf{6.8} & \textbf{3.3} & \textbf{7.3} \\
    \midrule
    \multicolumn{7}{l}{\textbf{100 hr labeled data}} \\
    wav2vec 2.0~\cite{baevski2020wav2vec} & 400k & 96 & 2.7 & 7.9 & 3.4 & 8.0 \\
    HuBERT~\cite{hsu2021hubert} & 250k + 400k & 47 & 2.7 & 7.8 & 3.4 & 8.1 \\
    WavLM~\cite{chen2022wavlm} & 250k + 400k & 187 & - & - & 3.4 & 7.7 \\
    data2vec~\cite{baevski2022data2vec} & 400k & 63 & 2.2 & 6.4 & 2.8 & 6.8 \\
    DinoSR~\cite{liu2024dinosr} & 400k & 63 & 2.3 & 6.4 & 2.9 & 6.7 \\
    data2vec 2.0~\cite{baevski2023efficient} & 400k & 17 & \underline{2.2} & \underline{6.2} & \underline{2.8} & \underline{6.4} \\ \cdashline{1-7}
     \method & 400k & 63 & \textbf{2.2} & \textbf{6.1} & \textbf{2.8} & \textbf{6.3} \\
\bottomrule
  \end{tabular}}
  \caption{\small Results on LibriLight benchmark and LibriSpeech for ASR. All the models share a similar BASE size encoder and are first fine-tuned with a 10 min / 1hr / 10hr / 100hr labeled dataset and then evaluated on common dev/test splits. The evaluation metric used is word error rate (WER). The best and the second best results are \textbf{bolded} and \underline{underlined} respectively}
  \label{tab:libri_light}
\end{table*}
\section{Results and Analysis}
In this section, we present the quantitative and qualitative results. For quantitative evaluation, we first fine-tune \method on LibriLight~\cite{kahn2020libri} and evaluate across all the test splits. Next, to show the scalability of the speech representations learned by \method, we conduct a downstream evaluation on SUPERB benchmark~\cite{yang2021superb}. Additionally, we also perform a qualitative analysis on the masked regions predicted by the \method. All the results reported for \method are averaged across five runs.      
\subsection{Evaluation on Low-Resource ASR}
For low-resource ASR evaluation, we follow a similar procedure as~\citet{baevski2020wav2vec} wherein we fine-tune only the student counterpart of \method with an additional CTC layer~\cite{graves2006connectionist} on top. We perform fine-tuning using low-resource labeled datasets under four different setups, 10min / 1hour / 10hour from LibriLight~\cite{kahn2020libri} and  100hour Librispeech~\cite{panayotov2015librispeech}.  
For evaluation, we use the standard dev/test split of Librispeech and report the word error rate (WER) by decoding with the official 4-gram language model. Following the prior work~\citet{baevski2022data2vec, baevski2023efficient}, the decoding hyper-parameter is searched with Ax (refer to Section~\ref{subsec:pre_fine}). As shown in Table~\ref{tab:libri_light}, \method consistently outperforms all the prior SSL methods across all the setups. We also provide additional results on other low-resource ASR benchmarks such as Wall Street Journal (WSJ)~\cite{paul1992design} and SwitchBoard (SB)~\cite{godfrey1992switchboard} in Appendix~\ref{sec:add_result}.
\subsection{Downstream Evaluation on SUPERB}
We extensively evaluate the effectiveness and scalability of the speech representation learned by \method using the Speech Processing Universal PERformance Benchmark (SUPERB). SUPERB, in total, consists of ten speech-related downstream tasks that aim to study four aspects of speech: content, speaker, semantics, and paralinguistics. To investigate the model's capabilities to understand speech content and semantics, we report the results on phoneme recognition (PR), automatic speech recognition (ASR), keyword spotting (KS), intent classification (IC), and slot filling (SF) (Additional details on all the downstream tasks can be found in Appendix~\ref{sec:superb}). For downstream evaluation on SUPERB, we follow a similar setup as~\citet{yang2021superb}, where we train a prediction head on top of the frozen pre-trained models instead of complete fine-tuning. As shown in Table~\ref{tab:superb_table}, for semantic tasks like IC and SF, the \method outperforms prior art, showing its capabilities to capture better semantic information from speech input. On context tasks, \method surpasses prior art in KS and achieves comparable performance on PR and ASR.
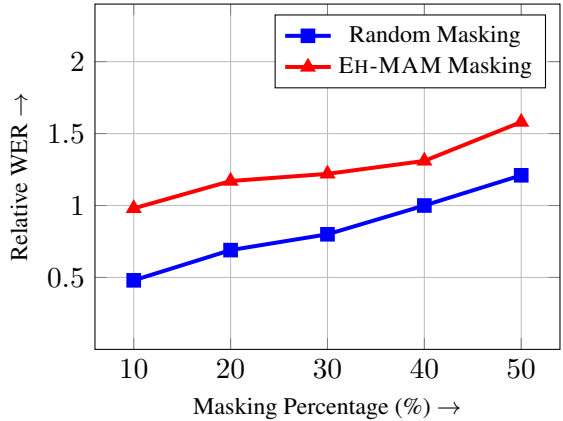
\begin{figure}
    \raggedright
    \begin{tikzpicture}
        \begin{axis}[
            xlabel={Masking Percentage (\%) $\rightarrow$},
            ylabel={Relative WER $\rightarrow$},
            legend pos=north east,
            grid=major,
            width=\columnwidth,
            height=0.8\columnwidth,
            ymin=0, ymax=2.4,
            xtick={10,20,30,40,50},
            ytick={0.5,1.0,1.5,2.0,2.5},
            legend style={font=\small},
            xlabel style={font=\small},
            ylabel style={font=\small},
        ]
        
        \addplot[
            color=blue,
            mark=square*,
            line width=1.5pt,
            mark options={solid, fill=blue},
            ]
            coordinates {
            (10, 0.48) (20, 0.69) (30, 0.8) (40, 1.0) (50, 1.21)
            };
        \addlegendentry{Random Masking}
        
        \addplot[
            color=red,
            mark=triangle*,
            line width=1.5pt,
            mark options={solid, fill=red},
            ]
            coordinates {
            (10, 0.98) (20, 1.17) (30, 1.22) (40, 1.31) (50, 1.58)
            };
        \addlegendentry{\method Masking}

        \end{axis}
    \end{tikzpicture}
    \caption{\small We compare the increase in relative Word Error Rate (WER) by selectively masking hard regions predicted by the loss predictor (\method Masking) Vs randomly masking frames. The increase in relative WER indicates that the \method Masking scheme masks useful context in an input.}
    \label{fig:ehmam_mask_comp}
\end{figure}
\subsection{Qualitative Analysis}
\noindent{\textbf{\method mask useful context:}} To show \method does mask useful context, we conduct a simple experiment wherein during ASR inference, we selectively mask the frames with high predicted reconstruction value using the loss predictor and compare the increase in relative WER with random masking. As shown in Fig.~\ref{fig:ehmam_mask_comp}, under SUPERB evaluation setting for ASR (refer Section~\ref{sec:exp}), we find selectively masking frames with \method constantly shows a higher relative WER when compared to random masking across various masking percentages. Higher relative WER indicates that a selective masking scheme with the \method masks useful context in a speech input.

\begin{figure}
    \raggedright
    \begin{tikzpicture}
        \begin{axis}[
            xlabel={Epochs $\rightarrow$},
            ylabel={Reconstruction Loss $\rightarrow$},
            legend pos=north east,
            grid=major,
            width=\columnwidth,
            height=0.8\columnwidth,
            ymin=0, ymax=15,
            xtick={20,40,60,80,100},
            ytick={3,6,9,12},
            legend style={font=\small},
            xlabel style={font=\small},
            ylabel style={font=\small},
        ]
        
        \addplot[
            color=green,
            line width=1.5pt,
            ]
            coordinates {
            (10, 11.48) (20, 10.69) (30, 9.46) (40, 8.16) (50, 7.56) (60, 6.45) (70, 5.0) (80, 4.6) (90, 4.2) (100, 3.7)
            };
        \addlegendentry{Hard Masking}
        
        \addplot[
            color=red,
            line width=1.5pt,
            ]
            coordinates {
            (10, 10.48) (20, 9.61) (30, 8.00) (40, 7.23) (50, 6.88) (60, 6.23) (70, 4.5) (80, 4.1) (90, 3.6) (100, 3.0)
            };
        \addlegendentry{\emph{easy-to-hard} Masking}

    \end{axis}
    \end{tikzpicture}
    \caption{\small We compare the effectiveness of \method in optimizing a MAM pretext task, such as the reduction in reconstruction loss, with hard and \emph{easy-to-hard} masking schemes. The \emph{easy-to-hard} masking scheme shows better convergence in reconstruction loss compared to hard masking strategies.}
    \label{fig:e2h_masking}
\end{figure}
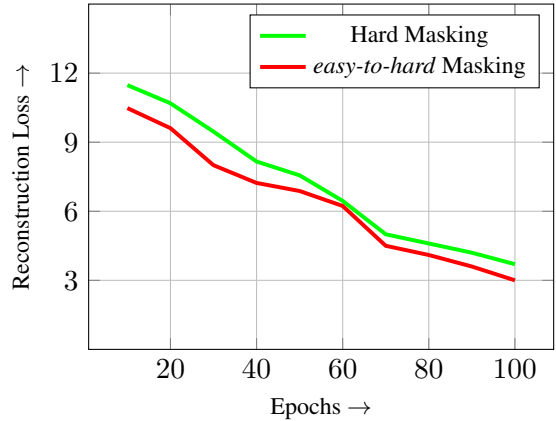

\noindent{\textbf{\method adapts well towards reconstructing hard regions:}} To show how well \method adapts towards reconstructing hard regions, we conduct an experiment wherein we compare the \method ability to reconstruct hard regions (collection of frames with high reconstruction values) using 1) \emph{hard masking}, masking only hard regions at each epoch and 2) \emph{easy-to-hard} masking, where we progressively introduce hard regions with randomly masked regions at each epoch. As shown in Fig.~\ref{fig:e2h_masking}, while pre-training \method, \emph{easy-to-hard} masking scheme shows better convergence in reconstruction loss when compared with hard masking strategies. This indicates that progressively introducing hard regions in an easy-to-hard manner, improves \method adaptability toward reconstructing masked regions during pre-training.

\section{Conclusion}
In this paper, we propose \method, a novel SSL framework for learning robust speech representations. In contrast to prior work that relies on random masking schemes for creating MAM pretext tasks, \method first identifies hard regions to reconstruct using a teacher network and then challenges the student to reconstruct them by progressively introducing hard regions throughout the learning process. Next, we introduce an \emph{easy-to-hard} masking scheme that guides the \method to mask harder regions to reconstruct step-by-step. \method outperforms all the other models on popular low-resource ASR benchmarks and downstream evaluation on SUPERB.

\section{Limitations and Future Work}
\method and our experimental setup have a few limitations, as mentioned below:
\begin{itemize}
    \item We do not employ a LARGE size encoder in \method, for example, a 24-layer variant used by~\citet{baevski2023efficient} due to compute constraints.
    \item The loss-predictors used in \method increase the trainable parameter count compared to other baselines such as data2vec 2.0~\cite{baevski2023efficient} during pre-training. However, we acknowledge that this accounts only for a slight increase in the total parameter count (roughly 5\%).
    \item Due to recourse constraints, we conduct the downstream evaluation on SUPERB for context and semantic-related tasks. We plan to extend the evaluation across speaker and paralinguistic tasks in the future. 
\end{itemize}

\section{Acknowledgements}
This project is supported in part by NSF\#1910940.

\bibliography{custom}
\appendix
\section{Baseline Details}
\label{sec:baseline_details}
{\noindent{\textbf{wav2vec 2.0.}~\footnote{\url{https://github.com/facebookresearch/fairseq/tree/main/examples/wav2vec}}}~\cite{baevski2020wav2vec} The wav2vec 2.0 model integrates contrastive learning with masking. Similar to the CPC model~\cite{oord2018representation}, it employs the InfoNCE loss~\cite{baevski2020wav2vec} to maximize the similarity between a contextualized representation (anchors) and a localized representation (positives) simultaneously minimizing the similarity with other masked regions (negatives). Instead of directly using the contextualized representations, wav2vec 2.0 employs a separate quantization module to generate positives and negatives.
\vspace{0.5mm}

{\noindent{\textbf{HuBERT.}}~\footnote{\url{https://github.com/facebookresearch/fairseq/tree/main/examples/hubert}}}}~\cite{hsu2021hubert} Like BERT~\cite{devlin-etal-2019-bert}, HuBERT follows a generative approach by discretizing the continuous MFCC features using the K-means algorithm and creating targets by randomly masking the quantized units. Unlike BERT, HuBERT employs a two-iteration training process wherein, in the first iteration, the model is trained to predict targets generated from the MFCC features, followed by quantizing the learned representations obtained from the first iteration training using K-means to generate new targets, which the model utilize in the second iteration training.
\vspace{0.5mm}

{\noindent{\textbf{WavLM.}}~\footnote{\url{https://huggingface.co/docs/transformers/en/model_doc/wavlm}}}~\cite{chen2022wavlm} WavLM extends the HuBERT's learning paradigm by introducing a gated relative position bias~\cite{chi2021xlm} at each transformer layer. Further, WavLM proposes an utterance-mixing strategy wherein training samples are augmented by mixing utterances from different speakers, and the targets are created from the original sample.
\vspace{0.5mm}

{\noindent{\textbf{data2vec.}}~\footnote{\url{https://github.com/facebookresearch/fairseq/tree/main/examples/data2vec}}}~\cite{baevski2022data2vec} data2vec introduces a self-distillation-based student-teacher networks for speech representation learning. The core idea is to predict the latent representations of the whole speech unlabeled data from the masked view. data2vec trains a student network by feeding a masked version of input to predict the latent representation obtained by feeding the whole input to a teacher network. The teacher's parameters are updated by the exponential moving average (ema) of the student's parameters.
\vspace{0.5mm}

{\noindent{\textbf{data2vec 2.0.}}~\footnote{\url{https://github.com/facebookresearch/fairseq/tree/main/examples/data2vec}}} ~\cite{baevski2023efficient} data2vec 2.0 uses an identical learning objective as data2vec but with two key changes. Firstly, data2vec 2.0 introduces a lightweight decoder module that reconstructs the masked frames in student representation before maximizing the similarity with the teacher representations. Next, data2vec 2.0 employs a multi-mask strategy where multiple mask variants of the same input are fed to the student network, followed by calculating reconstruction loss for all the variants with a common teacher representation obtained from the original speech input.
\vspace{0.5mm}

{\noindent{\textbf{DinoSR.}}~\footnote{\url{https://github.com/Alexander-H-Liu/dinosr}}}~\cite{liu2024dinosr} DinoSR uses similar architecture as~\cite{baevski2022data2vec} but introduces a novel gradient-free online clustering method for learning discrete acoustic units. DinoSR initially employs a teacher network to extract contextualized embeddings from the input audio. It then applies an online clustering scheme to these embeddings to create a machine-discovered phone inventory. Finally, it uses the discretized tokens to guide a student network.


\begin{table*}[h]
\centering
  \resizebox{0.8\textwidth}{!}{
  \begin{tabular}{{lccccccc}}
\toprule
\toprule
\textbf{Dataset} & \textbf{Language} & \textbf{Domain}  & \textbf{Type} & \begin{tabular}[c]{@{}c@{}}\textbf{Duration (hour)}\\ (train, dev, test)\end{tabular} \\ 
\midrule
LibriSpeech~\cite{panayotov2015librispeech} & English & General & Read & 960, 10, 10 \\
Libri-Light~\cite{kahn2020libri} & English & General & Read & 11.16, 10, 10 \\
SwitchBoard (SWBD)~\cite{godfrey1992switchboard} & English & Call Cent. & Conv. & 30, 5, N.A. \\
Wall Street Journal (WSJ)~\cite{paul1992design}  & English & Finance & Read & 80, 1.1, 0.4 \\ 
\bottomrule
  \end{tabular}}
  \caption{Detailed Statistics of datasets used in our low resource ASR evaluation. Type refers to Conversational or Read speech.}
  \label{asr_datasets}
\end{table*}
\begin{table*}[]
\centering
  \resizebox{\textwidth}{!}{
  \begin{tabular}{{lclccccc}}
\toprule
\toprule
\textbf{Task}                         & \textbf{Category} & \multicolumn{1}{c}{\textbf{Dataset}}                & \begin{tabular}[c]{@{}c@{}}\textbf{Duration(hour)}\\ \textbf{(train, dev, test)}\end{tabular} & \textbf{Evaluation Metric}\\ \midrule
Phoneme Recognition(PR)      & Content  & LibriSpeech\cite{panayotov2015librispeech}            & 100, 5.4, 5.4        &  Phoneme Error Rate(PER)                              \\
Automatic Speech Recognition(ASR) & Content  & LibriSpeech\cite{panayotov2015librispeech}            & 100, 5.4, 5.4     & Word Error Rate(WER)                                           \\
Keyword Spotting(KS)  & Content  & Speech Commands v0.1~\footnotemark\cite{warden2018speech}  & 18, 2, 1             & Accuracy(Acc)                           \\
Intent Classification(IC)  & Speaker  & Fluent Speech Commands~\footnotemark\cite{lugosch2019speech} & 23.1, 3.2, 3.9        &   Accuracy(Acc)                                     \\
Slot Filling(SF)  & Speaker  & Audio SNIPS~\footnotemark\cite{coucke2018snips}           & 166.0, 9.0, 9.0   & Concept Error Rate(CER)\\                      \bottomrule
  \end{tabular}}
  \caption{Details on downstream tasks and datasets used for the SUPERB evaluation}
  \label{dataset_details}
\end{table*}

\section{Dataset Details}
\label{sec:dataset_details}

\subsection{ASR Evaluation}
\label{sec:asr_evaluation}

{\noindent{\textbf{LibriSpeech.}}~\footnote{\url{https://www.openslr.org/12r}}}~\cite{panayotov2015librispeech} The LibriSpeech dataset is a widely-used corpus of English read speech with approximately 1000 hours of audiobooks available in the public domain, which includes a broad range of speakers, both male and female, with diverse accents and ages, providing a rich source for speech and language research. We pre-train our model on 960 hours LibriSpeech unlabeled data and fine-tune for ASR evaluation on 100 hours of labeled data.
\vspace{0.5mm}

{\noindent{\textbf{Libri-Light.}~\footnote{\url{https://github.com/facebookresearch/libri-light}}}~\cite{kahn2020libri}} The Libri-light is a dataset derived from the LibriVox project, consisting of audiobooks in the public domain, much like the LibriSpeech dataset, but aims to address the limitations of traditional ASR datasets by providing 60 hours of unlabelled speech complemented with a smaller amount of labeled data. We conduct evaluation on ASR with labeled data of 10 mins / 1 hour / 10 hours split from LibriLight.
\vspace{0.5mm}

{\noindent{ \textbf{WSJ.}~\footnote{\url{https://catalog.ldc.upenn.edu/LDC93S6A}}}~\cite{paul1992design}} The WSJ dataset consists of approximately 80 hours of read speech derived from articles in the Wall Street Journal, offering high-quality audio and transcriptions ideal for training and evaluating ASR systems. The WSJ dataset includes recordings from 84 speakers, providing diverse voice samples, including accurate word-level transcriptions for all audio files and metadata for speaker identities and recording conditions. We use the WSJ dataset for our ASR task evaluation on 80 hours of unlabeled data for training and 1.5 hours for of labeled data for testing.
\vspace{0.5mm}

{\noindent{\textbf{Switchboard.}~\footnote{\url{https://catalog.ldc.upenn.edu/LDC97S62}}}~\cite{godfrey1992switchboard}} The Switchboard is a telephone speech corpus consisting of approximately 260 hours of speech, which includes 2,400 two-sided telephone conversations among 543 speakers. The dataset conversations cover 70 different topics, from current events to personal interests, providing varied and natural discourse, making it an invaluable resource in the field of speech recognition, dialogue systems, and conversational analysis. We report the our ASR evaluation on 30 hours of unlabeled data for training and 5 hours of data for testing.

\footnotetext[7]{\url{https://www.tensorflow.org/datasets/catalog/speech_commands}}
\footnotetext[8]
{\url{https://fluent.ai/}}
\footnotetext[9]
{\url{https://github.com/aws-samples/aws-lex-noisy-spoken-language-understanding}}

\begin{table*}[t!]
\centering
\small
\begin{tabular}{lr}
\toprule \toprule
& Base (Librispeech) \\
\midrule
GPUs & 4  \\
Learning rate & ${7.5 \times 10^{-4}}$ \\
Adam $\beta_1$ / $\beta_2$ & 0.9 / 0.98 \\
Weight decay & 0.01  \\
Clip norm & - \\
Learning rate schedule & cosine \\
Warmup updates & 8,000  \\
Batch size & 63 min \\
$\tau_0$ (EMA start) & 0.999 \\ 
$\tau_e$ (EMA end) & 0.99999  \\ 
$\tau_n$ (EMA anneal steps) & 75,000  \\ 
$B$ (block width) & 5  \\
$R$ (mask ratio) & 0.5  \\
$A$ (mask adjust) & 0.05  \\
$K$ (layers to average) & 8 \\
Target normalization & IN $\rightarrow$ AVG \\
Updates & 400,000  \\
Decoder dim. & 384  \\
Decoder conv. groups & 16  \\
Decoder kernel & 7  \\
Decoder layers ($D$) & 4  \\
Loss Predictor dim. & 384  \\
Loss Predictor conv. groups & 16  \\
Loss Predictor kernel & 7  \\
Loss Predictor layers ($D$) & 4  \\
\bottomrule
\end{tabular}
\quad
\resizebox{0.5\textwidth}{!}{
\begin{tabular}{@{}lrrrr@{}}
\toprule \toprule
& 10 Minutes  & 1 Hour     & 10 Hours    & 100 Hours \\ \midrule
GPU                                                  & 4          & 4          & 4          & 4          \\
Learning rate                                       & ${5.0 \times 10^{-5}}$    & ${5.0 \times 10^{-5}}$    & ${5.0 \times 10^{-5}}$    & ${3.0 \times 10^{-5}}$    \\
Adam $\beta_1$ / $\beta_2$ & 0.9/0.98   & 0.9/0.98   & 0.9/0.98   & 0.9/0.98   \\
Learning rate schedule                               & tri\_stage & tri\_stage & tri\_stage & tri\_stage \\
Batch Size                                           & 32         & 32         & 32         & 32         \\
Updates                                              & 13000      & 13000      & 20000      & 80000      \\
Apply mask                                           & true       & true       & true       & true       \\
Mask prob                                           & 0.65       & 0.65       & 0.65       & 0.65       \\
Mask channel prob                                  & 0.25       & 0.25       & 0.50       & 0.50       \\
Mask channel length                                & 64         & 64         & 64         & 64         \\
Layerdrop                                            & 0.1        & 0.1        & 0.05       & 0.1        \\
Activation dropout                                  & 0.1        & 0.1        & 0.1        & 0.1        \\
Freeze finetune updates                              & 10000      & 10000      & 10000      & 0          \\ \bottomrule
\end{tabular}}
\caption{\small \textbf{(Left)} \method pre-training hyper-parameters. IN is instance normalization; AVG is mean pooling. \textbf{(Right)}  EH-MAM fine-tuning hyper-parameters for LibriLight \cite{kahn2020libri}}
\label{tab:finetune_hparams}
\end{table*}

\subsection{SUPERB (Speech processing Universal PERformance Benchmark)}
\label{sec:superb}

 SUPERB~\cite{yang2021superb} is a leaderboard to benchmark the performance of a shared model across a wide range of speech processing tasks with minimal architecture changes and labeled data. The key focus here is to extract the representation learned from SSL and to learn task-specialized lightweight
prediction heads on top of the frozen shared models. Below we detail the tasks in SUPERB that we use for evaluation.

\noindent\textbf{Phoneme Recognition (PR)} converts spoken language into its smallest units of sound, known as phonemes. This task incorporates alignment modeling to circumvent issues with incorrect forced alignments. The LibriSpeech ~\cite{panayotov2015librispeech} subsets train-clean-100/dev-clean/test-clean are utilized for training, validation, and testing in the SUPERB framework. The primary metric for evaluation is the phone error rate (PER).

\noindent\textbf{Automatic Speech Recognition (ASR)} transcribes spoken words into text. While PR focuses on the precision of phoneme modeling, ASR assesses improvements in terms of their practical relevance. The training, validation, and testing phases use the LibriSpeech ~\cite{kahn2020libri} subsets train-clean-100/dev-clean/test-clean. The word error rate (WER) serves as the evaluation metric.

\noindent\textbf{Keyword Spotting (KS)} involves the detection of specified keywords within speech, categorizing utterances into a set list of terms. The Speech Commands dataset v1.0 ~\cite{warden2018speech}, which includes ten keyword categories, a silence category, and an "unknown" category for erroneous detections, is used in this task. Accuracy (ACC) is the metric for assessing performance.

\noindent\textbf{Intent Classification (IC)} assigns categories to spoken utterances to ascertain the speaker's intent. It employs the Fluent Speech Commands ~\cite{lugosch2019speech} dataset, where utterances are labeled according to three intent categories: action, object, and location. The evaluation metric here is also accuracy (ACC).

\noindent\textbf{Slot Filling (SF)} entails predicting a series of semantic slots from speech. The Audio SNIPS ~\cite{coucke2018snips} dataset, which features synthesized multi-speaker utterances for the SNIPS NLU benchmark, is used for this purpose. Evaluation is based on the slot-type F1 score and slot-value character error rate (CER).

\section{Additional Details: Hyper-Parameter Tuning}
\subsection{Pre-training and Fine-tuning}
\label{subsec:pre_fine}
Table~\ref{tab:finetune_hparams} summarize the hyper-parameter choices for \method when pre-training on Librispeech-960 hours~\cite{panayotov2015librispeech} and fine-tuning across various LibriLight~\cite{kahn2020libri} setups (10min / 1hour / 10hour). Most hyper-parameters are taken from the prior art~\cite{baevski2023efficient, baevski2022data2vec}. For decoding, the hyper-parameter is searched using Ax~\footnote{\url{https://github.com/facebook/Ax}}
\begin{table}[h]
    \centering
    \resizebox{0.75\columnwidth}{!}{
    \begin{tabular}{@{}c|lllll@{}} 
    \hline \hline
$\alpha$     & 1 & 0.5 & 0.1 & 0.05 & 0.01 \\ \hline
WER $\downarrow$   & 7.4 & 7.3  & 7.3  & \textbf{7.1}  & \underline{7.2} \\ \hline
    \end{tabular}}
    \caption{\small $\alpha$ = 0.05 gives the best performance}
    \label{tab:change_alpha}
\end{table}
\subsection{Balancing Parameter $\alpha$}
\label{subsec:bal_par}
In Table~\ref{tab:change_alpha} we show the effect of changing the balancing parameter $\alpha$, on the final low-resource asr evaluation. Specifically, we pre-train \method with different balancing parameters and then perform an end-to-end fine-tuning on 10 mins setup of LibriLight~\cite{kahn2020libri}. Finally, we compute WER for the test-clean split.

\subsection{Masking Probability $\mathcal{P}$}
\label{subsec:mask_prob}
In Table~\ref{tab:mask_p} we show the effect of changing the masking probability $\mathcal{P}$, on the final low-resource asr evaluation. Specifically, we pre-train \method with different masking probability and then perform an end-to-end fine-tuning on 10 mins setup of LibriLight~\cite{kahn2020libri}. Finally, we compute WER for the test-clean split.

\begin{table}[h]
    \centering
    \resizebox{0.75\columnwidth}{!}{
    \begin{tabular}{@{}c|lllll@{}} 
    \hline \hline
$\mathcal{P}$     & 0.1 & 0.2 & 0.3 & 0.4 & 0.5 \\ \hline
WER $\downarrow$   & 9.4 & 8.1  & 7.9  & \underline{7.3}  & \textbf{7.1} \\ \hline
    \end{tabular}}
    \caption{\small $\mathcal{P}$ = 0.5 gives the best performance}
    \label{tab:mask_p}
\end{table}
\begin{table}[h]
\centering
\resizebox{\columnwidth}{!}{
\begin{tabular}{cccc}
\toprule \toprule
Models & \multicolumn{2}{c}{WSJ (WER $\downarrow$)}       & Switchboard (WER $\downarrow$)\\ \cmidrule{2-4}
       & dev & test & dev        \\ \midrule
data2vec 2.0     & 9.2   &   9.0   &     15.2         \\ \cdashline{1-4} 
\method       &   \textbf{8.4}  &  \textbf{8.2}    &       \textbf{14.3}       \\\bottomrule
\end{tabular}}
\caption{\small Performance comparison of \method on WSJ and Switchnoard datasets.}
\label{tab:add_reslts_tab}
\end{table}
\label{sec:hyper_parametes}

\section{Additional Details: General}
\label{sec:gen}
{\noindent{\textbf{Compute details.}}} For all our pre-training and fine-tuning experiments, we used four NVIDIA A100-40GB GPUs. The pre-training \method requires five days of training and consists of 94.40M parameters. All the fine-tuning experiments on LibriLight~\cite{kahn2020libri} require two days each. Additionally, individual downstream evaluation on SUPERB requires one day.
\vspace{0.5mm}

{\noindent{\textbf{Potential Risk.}}} As the \method follows a self-supervised training regime, it may learn spurious correlations which can affect downstream performance on ASR, PR, etc. Moreover, \method might get biased towards a particular type of accent, dialect, or domain, such as telephonic or read speech, due to a huge amount of unlabeled data, which may not be diverse.

{\noindent \textbf{Software and Packages details.}} We implement all our models in PyTorch ~\footnote{\url{https://pytorch.org/}} and use Fairseq~\footnote{\url{https://github.com/facebookresearch/fairseq}} toolkit and SUPERB~\footnote{\url{https://superbbenchmark.github.io}} for all our experiments.

\section{Additional Results}
\label{sec:add_result}
We present additional results for low-resource ASR evaluation on WSJ~\cite{paul1992design} and Switchboard~\cite{godfrey1992switchboard}. The evaluation settings for both datasets are similar to LibriLight~\cite{kahn2020libri}. Train/Test splits for both datasets can be found in Table~\ref{asr_datasets}. As shown in Table~\ref{tab:add_reslts_tab}, \method outperforms the state-of-the-art model, data2vec 2.0, across all the dev/test splits.  
\label{sec:add_results}

\begin{algorithm*}[t]
\caption{Pseudo-Code of Easy-to-Hard Masking.}
\label{alg:mask}
\tiny
\begin{lstlisting}[language=python]
def compute_mask_indices_ema_loss(
    shape: Tuple[int, int],
    padding_mask: Optional[torch.Tensor],
    loss_pred: Optional[torch.Tensor],
    mask_prob: float,
    mask_length: int,
    current_epoch: int,
    total_epoch: int,
    mask_type: str = "static",
    min_masks: int = 0,
    require_same_masks: bool = True,
    mask_dropout: float = 0.0):
    
    bsz, all_sz = shape
    mask = np.full((bsz, all_sz), False)
    # add a random number for probabilistic rounding
    all_num_mask = int( mask_prob * all_sz / float(mask_length) + np.random.rand())
    # Get the loss lattice from decoder
    ids_shuffle_loss = torch.argsort(loss_pred, dim=1).cpu().detach().numpy()

    all_num_mask = max(min_masks, all_num_mask)
    # guide the making wrt to training epoch
    #keep_ratio = 1.0
    keep_ratio = float((current_epoch + 1) / total_epoch)
    mask_idcs = []
    for i in range(bsz):
        if padding_mask is not None:
            sz = all_sz - padding_mask[i].long().sum().item()
            num_mask = int(
                # add a random number for probabilistic rounding
                mask_prob * sz / float(mask_length)
                + np.random.rand()
            )
            num_mask = max(min_masks, num_mask)
        else:
            sz = all_sz
            num_mask = all_num_mask

        if mask_type == "static":
            lengths = np.full(num_mask, mask_length)
        else:
            raise Exception("unknown mask selection " + mask_type)

        if sum(lengths) == 0:
            lengths[0] = min(mask_length, sz - 1)

        min_len = min(lengths)
        if sz - min_len <= num_mask:
            min_len = sz - num_mask - 1
        #reverse the index list to get the indexes associated with max losses
        sample_loss_index = ids_shuffle_loss[i][::-1]
        
        #calculate random_mask and learnable_mask using keep_ratio
        random_mask = int(num_mask * (1-keep_ratio))
        learnable_mask = num_mask - random_mask
        
        #randomly select mask index. 
        mask_idc = np.random.choice(sz - min_len, num_mask, replace=False)
        sample_loss_index = sample_loss_index[:learnable_mask]
        
        #recalculate mask_idc for random masking:
        mask_idc = np.random.choice(np.setdiff1d(mask_idc, sample_loss_index), random_mask, replace=False)

        loss_mask_idc = np.asarray(
            [
                sample_loss_index[j] + offset
                for j in range(len(sample_loss_index))
                for offset in range(lengths[j])
            ]
        )
        
        mask_idc = np.asarray(
            [
                mask_idc[j] + offset
                for j in range(len(mask_idc))
                for offset in range(lengths[j])
            ]
        )
        #print(loss_mask_idc)
        
        if len(mask_idc) == 0:
            combine_idc = loss_mask_idc
        else:
            combine_idc = np.concatenate((loss_mask_idc, mask_idc))
        
        mask_idcs.append(np.unique(combine_idc[combine_idc < sz]))
    min_len = min([len(m) for m in mask_idcs])
    for i, mask_idc in enumerate(mask_idcs):
        if len(mask_idc) > min_len and require_same_masks:
            mask_idc = np.random.choice(mask_idc, min_len, replace=False)
        if mask_dropout > 0:
            num_holes = np.rint(len(mask_idc) * mask_dropout).astype(int)
            mask_idc = np.random.choice(
                mask_idc, len(mask_idc) - num_holes, replace=False
            )
        mask[i, mask_idc] = True
    return mask
\end{lstlisting}
\end{algorithm*}

\end{document}